\documentstyle[PASJadd,epsf]{PASJ95}
\newcommand{\vol}{}
\newcommand{\no}{}
\newcommand{\lett}{}
\newcommand{\spage}{}
\newcommand{\rdate}{}
\newcommand{\adate}{}

\begin{document}
\setcounter{page}{1}

\title{ Galaxy Formation from
a Low-Spin Density Perturbation\\ in a CDM Universe}

\author{ Daisuke {\sc Kawata}
\\
{\it
Astronomical Institute, Tohoku University, Sendai, Miyagi 980-8578}
\\
{\it  E-mail(DK): kawata@astr.tohoku.ac.jp}
}

\abst{
In order to understand the formation process of
elliptical galaxies which are not rotationally supported,
we have carried out numerical simulations of
the galaxy formation from the density perturbation with
a rotation corresponding to a small spin parameter.
The three-dimensional TREE N-Body/SPH simulation code
used in this paper includes the dark matter and gas dynamics,
radiative cooling, star formation, supernova feedback, and
metal enrichment. The initial condition
is a slowly rotating, top-hat over-dense sphere
on which the perturbations expected in a cold dark matter (CDM) universe
are superposed. By means of the stellar population synthesis,
we calculated the surface brightness profile,
the metallicity distribution, and
the photometric properties of the end-product, and
found that these properties quantitatively agree with
the observed properties of bright elliptical galaxies.
Thus, we conclude that, in a CDM universe,
the proto-galaxy which has a spin-parameter as small as 0.02
evolves into an elliptical galaxy.
}
\kword{Galaxies: formation --- Galaxies: elliptical --- Numerical methods}
\maketitle
\thispagestyle{headings}

\section{Introduction}

The new generation of large ground-based telescopes
and the Hubble Space Telescope allow us to
observe structures and properties of galaxies
in unprecedented detail. In order to clarify the history of galaxy
formation and evolution based on the observed features of galaxies,
the numerical simulations are required
which can be compared with these observations directly.
The numerical models of galaxy formation have progressed greatly
since it was first developed by a series of Larson's papers.
Larson (1969) modeled
the gas dissipation and the chemical evolution in galaxy formation and 
reproduced main properties of the observed elliptical
(Larson 1974a, 1974b, 1975)
and disk (Larson 1976) galaxies successfully.
However, he did not take into account the details of the hydrodynamics and
dissipation of the gas, and only considered the axisymmetric system.
Carlberg (1984a, 1984b) and Carlberg et al.\ (1986)
simulated the dissipational formation of spheroidal galaxies
using N-body model. They succeeded in reproducing
elliptical galaxies with a de Vaucouleurs surface density profile.
Although they improved Larson's approach, some problems still remained.
For instance, the treatment of
the gas component and of the star formation process was phenomenological,
because they introduced various parameters, such as
the collision rate of the clouds,
the rates of dissipation and star formation
after encounter between gas particles, and the strength
of the kinetic feedback to the surrounding gas
caused by supernovae of new born stars.
Furthermore, their simulations were performed in dimensionless units,
so that there was no physically convincing way of scaling
their results of simulations to dimensional units and comparing
them with the observational data.

 Katz \& Gunn (1991) presented the modeling galaxy formation 
using the general-purpose code for evolving self-gravitating
fluids in three dimensions, called TreeSPH (Hernquist, Katz 1989;
Katz et al. 1996). TreeSPH is a fully Lagrangian code to combine
smoothed particle hydrodynamics (SPH; Lucy 1977; Gingold, Monaghan
1977) with a hierarchical tree algorithm for computations
of gravitational forces (Barnes, Hut 1986).
Since in this method the thermodynamic state of the gas
can be calculated, the dissipational effects of the gas
are included as the radiative cooling following the standard cooling curves.
Thus, simulations based on this numerical model can lead to physical values
which can be compared with the observational data quantitatively.
Furthermore, the TreeSPH simulation was presented which includes
the process of the star formation (Katz 1992) and
of metal enrichment due to supernova (Steinmetz, M\"uller 1994, 1995).

 Using these numerical model, they (Katz, Gunn 1992; Katz 1992;
Steinmetz, M\"uller 1994, 1995) investigated
the collapse of an isolated constant-density perturbation
which initially follows a Hubble-flow expansion 
and has a solid-body rotation, since external tidal fields
are not included.
The top-hat density perturbation consisted of
dark and baryonic matter, and included small scale perturbations assuming
power spectrum index $-2.5$.
In their model, the amount of initial solid-body rotation
was specified by the dimensionless spin parameter, $\lambda$
(Peebles 1971). They studied only the case with $\lambda = 0.08$,
and seemed to succeed in forming a system
which had some similar properties to an observed
disk galaxy. However, some groups (e.g., Barnes, Efstathiou 1987;
Warren et al.\ 1992) concluded that
the spin parameter of a virialized dark matter
in a cold dark matter (CDM) universe falls on a range of $0.02-0.11$ 
with the median value of $0.05$.  The recent analytical studies also
derived the distribution of spin parameter with
the similar spread and median value (e.g., Steinmetz, Bartelmann 1995;
Eisenstein, Loeb 1995), in which $\lambda=0.08$ is near a high end value.
By this reason, it is worth while investigating
the properties of the system formed 
from the initial density perturbation with a small value of $\lambda$.

 Therefore, we calculate the dynamical and chemical evolution
of density perturbations with the small spin parameter, $\lambda=0.02$.
The present simulation follows a semi-cosmological manner (Katz, Gunn 1991)
in order to specify initial density and velocity fields, 
and adopts the method of numerical simulation used by
Katz (1992) and Steinmetz \& M\"uller (1994, 1995).
Using stellar population synthesis (Kodama, Arimoto 1997),
we can directly compare the results of simulations with
the observed properties of galaxies, such as
the surface brightness profile,
the half-light radius, the metallicity gradient,
and the color gradient.

 According to Heavens \& Peacock's (1988) analytical estimation,
high density peak tends to have a small value of $\lambda$.
This reason is that higher peaks have a shorter collapse time;
thus they have less time to get spun up by the environmental tidal force.
On the other hand, bright elliptical galaxy is considered as the system
which collapsed at high redshift and has a small rotation.
It is a natural supposition that ellipticals were formed
from a high density peak (e.g., Blumenthal et al. 1984).
Hence, we focus on comparison the results of numerical simulation
with the observed properties of elliptical galaxies.

 Section 2 presents the method of numerical simulations, and
the details of the top-hat model of galaxy formation
are described in section 3.
 Section 4 presents the results of numerical simulations.
Summary and discussion are described in section 5.

\section{The Code}

 In our numerical simulation,
the dynamics of the dark mater and stars is
calculated by N-body scheme, and
the gas component is modeled using the SPH.
The code also includes the processes of the gas cooling
and of the star formation.
Many authors
(e.g., Hernquist, Katz 1989; Navarro, White 1993; Katz et al.\ 1996;
Steinmetz 1996; Carraro et al.\ 1998) have already described
the details of numerical model for galaxy formation
which employs the SPH method.
Our code is based on these previous modelings essentially
and is the revised version of the SPH code in Kawata, Hanami (1998)
for the application to galaxy formation. We describe details of
our SPH code and the model of the cooling and
star formation in the following.

\subsection{The SPH Code}

 In the SPH method, the densities and other local quantities in the gas
component are computed by a kernel estimation.
We employ a spherically symmetric spline kernel (Steinmetz 1996),
\begin{equation}
\begin{array}{l}
W(r, h) = \frac{8}{\pi h^{3}} \nonumber \\
 \times \left\{ \begin{array}{cc}
 1-6(r/h)^{2}+6(r/h)^{3} & {\rm if}\ 0\leq r/h\leq 1/2, \nonumber \\
 2[1-(r/h)]^{3}      & {\rm if}\ 1/2\leq r/h\leq 1, \nonumber \\
 0               & {\rm otherwise}.
\end{array} \right.\\
\end{array} 
\label{Weq}
\end{equation}
 Each particle has its own smoothing length, $h_i$, chosen so that
a fixed number of neighbor particles are contained within
its smoothing length (e.g., Hernquist, Katz 1989).
Using this kernel, the smoothed density associated with the $i$-th particle
is given by 
\begin{equation}
\rho_i = \sum_{j} m_j W(r_{ij}, h_{ij}),
\end{equation}
 where $r_{ij} \equiv |\mbox{\boldmath $x$}_i-\mbox{\boldmath $x$}_j|$ and
$h_{ij} = (h_i+h_j)/2$ is the pair-averaged smoothing length.

 The acceleration of the $i$-th particle is determined by Euler's equation,
\begin{equation}
 \frac{d \mbox{\boldmath $v$}_i}{dt} = 
 - \frac{1}{\rho_i} \nabla_i ( P + P_{\rm visc} ) - \nabla_i \Phi,
\label{Eueq} 
\end{equation}
 where $\Phi$, $P$, and $P_{{\rm visc}}$ are
the gravitational potential, the pressure, and
the artificial viscous pressure, respectively.
 The pressure-gradient term can be written as (e.g., Steinmetz 1996)
\begin{equation}
\begin{array}{l}
  \frac{1}{\rho_i} \nabla_i ( P + P_{\rm visc} ) \\
 = \sum_{j}
 m_{j} \left(
\frac{P_{i}}{\rho_{i}^{2}} + \frac{P_{j}}{\rho_{j}^2} + Q_{ij} \right)
 \nabla_{i} W(\mbox{\boldmath $x$}_{ij}, h_{ij}),
\end{array} 
\end{equation}
 where
\begin{equation}
 Q_{ij} =
\left\{
 \begin{array}{cc}
\displaystyle{\frac{ -\alpha v_{{\rm s},ij} \mu_{ij} + \beta \mu_{ij}^{2}}
{\rho_{ij}}} &
 {\rm if}\ \mbox{\boldmath $x$}_{ij}\cdot\mbox{\boldmath $v$}_{ij}<0, \\
0 & {\rm otherwise},
 \end{array}
\right.
\label{Qeq}
\end{equation}
with
\begin{equation}
 \mu_{ij} =
\frac{ 0.5 h_{ij}\mbox{\boldmath $v$}_{ij}
\cdot \mbox{\boldmath $x$}_{ij}}
{r_{ij}^{2} + \eta^{2}}.
\label{Myuijeq}
\end{equation}
Here, we define
$\mbox{\boldmath $x$}_{ij}
\equiv\mbox{\boldmath $x$}_i-\mbox{\boldmath $x$}_j$
and $\mbox{\boldmath $v$}_{ij}
\equiv\mbox{\boldmath $v$}_i-\mbox{\boldmath $v$}_j$.
In this paper,
we chose parameters $(\alpha, \beta)=(0.5, 1.0)$ in equation (\ref{Qeq})
and $\eta=0.05 h_{ij}$ in equation (\ref{Myuijeq}) so that the features
of the one-dimensional shock tube can be well reproduced.
We use the corrected artificial
viscosity, which was proposed by Navarro \& Steinmetz (1997): 
\begin{eqnarray}
 \tilde{Q}_{ij} & = & Q_{ij}\frac{f_i+f_j}{2}, \nonumber \\
  f_i & = &
  \frac{|\langle {\bf \nabla}\cdot \mbox{\boldmath $v$}\rangle_i|}
       {|\langle {\bf \nabla}\cdot \mbox{\boldmath $v$}\rangle_i|
       +|\langle {\bf \nabla}\times\mbox{\boldmath $v$}\rangle_i|
       +0.0002 v_{{\rm s},i}/h_i},
\label{shearv}
\end{eqnarray}
where $v_{\rm s}$ is the sound velocity. The velocity divergence,
$\langle {\bf \nabla} \cdot \mbox{\boldmath$v$}\rangle_i$,
and rotation, $\langle {\bf \nabla}\times\mbox{\boldmath $v$}
\rangle_{i,x}$, of the $i$-th particle
are calculated by
\begin{equation}
 \langle\nabla \cdot \mbox{\boldmath $v$} \rangle_i
  = -\frac{1}{\rho_i}
 \sum_{j} m_j \mbox{\boldmath $v$}_{ij}
 \cdot \nabla_i W(\mbox{\boldmath $x$}_{ij}, h_{ij}), 
\end{equation} 
%
\begin{equation}
\begin{array}{l}
\langle \nabla \times \mbox{\boldmath $v$} \rangle_{i, x}
 = -\frac{1}{\rho_i} \sum_{j} m_j \\
\times  \left[v_{ij,z}
 \nabla_{i,y} W(\mbox{\boldmath $x$}_{ij},h_{ij})
- v_{ij,y}
 \nabla_{i,z} W(\mbox{\boldmath $x$}_{ij},h_{ij}) \right],
\end{array} 
\end{equation}
where
$v_{ij,x}\equiv v_{i,x}- v_{j,x}$.
 The gravitational-force term of the $i$-th particle
is the summation of the contributions from dark matter, gas,
and star particles,
\begin{eqnarray}
 \nabla_i \Phi & = &
- G \sum_{j} \frac{m_{j}\mbox{\boldmath $x$}_{ij}}
{(r_{ij}^2+\varepsilon_{ij}^2)^{3/2}},
\label{grav}
\end{eqnarray}
where $\varepsilon_{ij}\equiv(\varepsilon_i+\varepsilon_j)/2$
is the mean softening length of $i$ and $j$-th particles.
The gravitational force for each particle is computed
using the tree method (e.g., Barnes, Hut 1986; Pealzner, Gibbon 1996)
with expansions to the quadrupole order and
the tolerance parameter $\theta=0.8$.

 The evolution of the internal energy, $u_i$, of the $i$-th particle is
determined by the energy equation (e.g., Steinmetz 1996),
\begin{equation}
\begin{array}{l}
\frac{d u_i}{dt} = \frac{P_i}{\rho_i} \sum_{j} m_j
  \mbox{\boldmath $v$}_{ij} \cdot
\nabla_i W(\mbox{\boldmath $x$}_{ij}, h_{ij})\\
 +\frac{1}{2} \sum_{j} m_j Q_{ij}
 \mbox{\boldmath $v$}_{ij} \cdot
\nabla_i W(\mbox{\boldmath $x$}_{ij}, h_{ij})
 -\frac{\Lambda_i}{\rho_i}+{\cal H}_i,
\label{Eneeq}
\end{array}
\end{equation}
where $\Lambda /\rho$ and ${\cal H}$ are
the cooling rate and the heating rate per unit mass,
respectively.
In this paper, we consider the cooling which is due to the radiative process
of the gas having a specified metallicity
and the heating which is caused by the feedback of the massive stars
(see following sections for detail).
The pressure, $P$, of each particle
can be obtained by the equation of state for an ideal gas,
\begin{equation}
 P = (\gamma-1) \rho u,
\end{equation}
where $\gamma=5/3$ for a mono-atomic gas.

 The evolution of the smoothing length, $h$, is
determined so that each particle has 
a roughly constant number of neighbors, i.e.,
the particles located within $h$ from that particle.
Thus, the smoothing length, $h^{n+1}$, at a time step $n+1$
is determined by the smoothing length, $h^{n}$,
and the number of neighbors, $N^{ n}$, at the previous time step $n$;
\begin{equation}
 h^{n+1} = h^{n}\frac{1}{2}
 \left[1+\left(\frac{N_{\rm s}}{N^{n}}\right)^{1/3}\right],
\end{equation}
where $N_{\rm s}$ is a parameter (Hernquist, Katz, 1989).
Our one-dimensional shock-tube tests have revealed that
the acceptable range of $N_{\rm s}$ is $[34, 68]$ in the
three dimensional space. In this paper,
we adopt $N_{\rm s}=40$.

 In order to update the velocity and the position of
each particle, equation (\ref{Eueq}) is
integrated using a leap-frog algorithm with individual time steps
(Hernquist, Katz 1989; Makino 1991).
The time step of the $i$-th particle is chosen to be
 $\Delta t_{i} = \min(0.3 \Delta t_{{\rm CFL},i},
 0.1 \Delta t_{{\rm f},i})$,
where $\Delta t_{{\rm CFL}}$
is determined by the Courant--Friedrichs--Lewy condition as
\begin{equation}
\Delta t_{{\rm CFL},i} =
 \frac{0.5 h_i}{v_{{\rm s},i}+1.2(\alpha v_{{\rm s},i}+ \beta
{\rm max}_j|\mu_{ij}|)},
\label{Couranteq}
\end{equation}
and  $\Delta t_{{\rm f}}$ is determined by the requirement
that the force should not
change too much from one time step to the next time step,
leading to the specification
\begin{equation}
 \Delta t_{{\rm f},i} = \sqrt{\frac{h_i}{2}
  \left|\frac{d \mbox{\boldmath $v$}_i}{dt}\right|^{-1}}.
\label{acceeq}
\end{equation}
In a dense region, the smoothing length may become 
shorter than the softening length, requiring a very small time step.
In order to avoid this,
we set the lower limit of the smoothing length
to be $h_{\rm min}=\varepsilon/2$.
 For the collisionless particles which represent dark matter and stars,
the time step is determined by 
$\Delta t = \min[0.16 (\varepsilon/ v),
 0.4 (\varepsilon/ |dv/dt|)^{1/2}]$.

 These time steps do not always satisfy
the constraints from the time scale of the internal
energy evolution, because the time scale of the radiative cooling
can be very small. Thus, when the evolution of
the internal energy of each particle is integrated,
we do not use the individual time step scheme but
the integration is carried out every time step which is the minimum
among all the particles. 
Moreover, the energy equation is integrated using
a semi-implicit method (Hernquist, Katz 1989).
In the implementation, the radiative cooling is damped by
the following equation
in order to ensure numerical stability (Katz, Gunn 1991),
\begin{eqnarray}
\left(\frac{du}{dt}\right)_{{\rm rad}}^{\rm damped}
&=& \frac{a (du/dt)_{\rm rad}}{\sqrt{a^2+[(du/dt)_{\rm rad}]^2}},\nonumber\\
a&=&0.5\frac{u}{\Delta t}+\left(\frac{du}{dt}\right)_{\rm ad}.
\end{eqnarray}
Here, $(du/dt)_{\rm ad}$ is the change in the thermal energy
due to the adiabatic compression or expansion of the gas
excluding the contribution from the undamped
radiative cooling, $(du/dt)_{\rm rad}$.

 Our code has been vectorized and parallelized using
the method of Makino (1990).
The parallelization is done only in calculating
gravitational forces and searching neighbors.
Since these calculations occupy most of the computational time,
this parallelization is very beneficial.
The MPI library allows our code to run on any computers from the serial
to the parallel one, and we can obtain the same results on different computers.

\subsection{The Radiative Cooling}

 We employ the cooling model in Theis et al.\ (1992)
who approximated the cooling rate
calculated by Dalgarno \& McCray (1972) and
B\"ohringer \& Hensler(1989) as follows;
\begin{equation}
\Lambda(T,Z,n_{\rm H}) = \Lambda_0(Z)T^{m(Z)} n_{\rm H}^2
\ \ {\rm erg\ cm^{-3}\ s^{-1}},
\end{equation}
where $Z$ and $n_{\rm H}$ are the metallicity
and the hydrogen number density of the gas, respectively.
$\Lambda_0(Z)$ and $m(Z)$ are defined for different temperature ranges
as follows;
\\
at $10^4\ {\rm K} \leq T < 10^5\ {\rm K}$,
\begin{eqnarray}
m(Z) & = & 30 Z, \nonumber \\
\Lambda_0(Z) & = & 4 \times 10^{-22.1-4 m(Z)}, \nonumber
\end{eqnarray}
at $T \geq 10^5\ {\rm K}$,
\begin{eqnarray}
m(Z) & = & \frac{2.5+7 \sqrt{Z}}
{5-\log (1.48 \times 10^{11} Z^{1.1}+10^6)}, \nonumber \\
\Lambda_0(Z) & = & 10^{-22.1-5 m(Z)+7 \sqrt{Z}}. \nonumber
\end{eqnarray}
 For temperatures above $5 \times 10^4$ K, bremsstrahlung becomes
important, which is expressed here as:
\begin{equation}
 \Lambda_{\rm br} = 1.4 \times 10^{-27} (f^{\rm ion})^2
 n_{\rm e} n_{\rm ion} T^{0.5} \langle g_{\rm ff} \rangle,
\end{equation}
 where $\langle g_{\rm ff} \rangle \approx 1.35$ is an averaged
Gaunt factor and,
\begin{eqnarray}
 f^{\rm ion} & = & n_{\rm e}/ n_{\rm ion} \nonumber,\\
 n_{\rm ion} & = &
 \frac{n_{\rm H}}{1-Y-Z}\left(1-\frac{3}{4}Y-\frac{17}{18}Z\right),
\nonumber \\ 
 n_{\rm e} & = &
 \frac{n_{\rm H}}{1-Y-Z}\left(1-\frac{1}{2}Y-\frac{7}{9}Z\right).
\nonumber 
\end{eqnarray}
Here, $Y$ denotes the mass fraction of helium and
we set $Y=0.24$. 
We do not consider the cooling below $10^4$ K, because our simulation
does not have enough mass resolution
for cooling processes in such low temperatures
(e.g., ${\rm H}_{\rm 2}$ cooling) and because the cooling rate declines
rapidly below $10^4$ K. Furthermore, the heat sources such as
the UV background radiation, which are ignored in this paper, could warm
the gas components to this temperature easily.
Thus, we set the lower limit of the temperature to be
$T_{\rm lim}=10^4\ {\rm K}$ (Katz, Gunn 1991).

\subsection{The Star Formation}

 We model the star formation using a method similar to
that of Katz (1992) and Katz et al.\ (1996).
We assume that the star formation occurs
where the gas density is larger than
a critical density,
$\rho_{\rm crit} = 7 \times 10^{-26}\ {\rm g\ cm^{-3}}$,
and where the gas velocity field is convergent,
${\bf \nabla} \cdot \mbox{\boldmath $v$}_i < 0$.
 When a gas particle is eligible to form stars,
its star formation rate (SFR) is,
\begin{equation}
 \frac{d \rho_*}{dt} = -\frac{d \rho_{g}}{dt}
 = \frac{c_* \rho_g}{t_g},
\label{sfreq}
\end{equation}
where $c_*$ is a dimensionless SFR parameter
and $t_g = \sqrt{3 \pi/16 G \rho_g}$ is the
dynamical time, which is longer than the cooling time scale
in the region eligible to form stars.
This formula corresponds to the Schmidt law that
SFR is proportional to $\rho_{g}^{1.5}$.
In this paper, we set $c_* = 1$.
 Equation (\ref{sfreq}) implies that the probability, $p_*$,
which one gas particle forms stars during
a discrete time step, $\Delta t$, is
\begin{equation}
 p_* = 1 - \exp (-c_* \Delta t /t_g).
\end{equation}
In order to avoid making an untolerably large number of new star particles
with different masses, we set the whole gas particle 
transforming to one star particle with this probability.
After the particle changed to the star,
it behaves as a collisionless particle dynamically.

 We take into account the energy feedback to the surrounding gas by supernovae.
We assume that each supernova yields thermal energy of $10^{51}$ ergs.
For simplification,  we assume that
each massive star ($>8 {M_\odot}$)
explodes within the simulation time step in which it was born
(instantaneous recycling).
In our simulation, the feedback also adds the ejected mass
to the surrounding gas. In this process, a portion of metals
produced in the stars is also returned to the gas,
leading to chemical enrichment. We assume that
an $m$ ${M_\odot}$ star yields $(0.357m-2.2)$ ${M_\odot}$
of metal (Maeder 1987; Steinmetz, M\"uller 1994).
The ejected mass is $(m-1.4)$ ${M_\odot}$, where the remnant mass
is assumed to be $1.4\ {M_\odot}$.
The feedback energy, ${E_{\rm SN}}$,
the ejected mass, ${M_{\rm SN}}$, and the ejected metals,
 ${M{\rm z}_{\rm SN}}$, can be obtained from the Salpeter IMF ($x=1.35$) with
the lower mass limit of 0.1 ${M_\odot}$ and
the upper mass limit of 60 ${M_\odot}$.
This IMF leads to
${M_{\rm SN}}\sim 0.122\ {M_\odot}$,
${E_{\rm SN} \sim 0.007310\times 10^{51}\ \rm ergs}$, and
${M{\rm z}_{\rm SN} \sim 0.02754\ M_\odot}$,
for each $1\ {M_\odot}$ star born.
These mass, energy, and metals 
are smoothed over the neighbor particles using the SPH
smoothing algorithm.
For example, when the $i$-th particle changes from the gas to a star,
the increment of the mass of the $j$-th neighbor particle
due to explosion of the new star is given by
\begin{equation}
 \Delta { M_{\rm SN,{\it j}}}
  =  \frac{m_j}{\rho_{g,i}} { M_{\rm SN,{\it i}}},
\end{equation}
where
\begin{equation}
 \rho_{g,i} = \langle \rho_{g}(\mbox{\boldmath $x$}_i) \rangle
 = \sum_{j \neq i} m_j W(r_{ij},h_{ij}).
\end{equation}

\section{The Model}

Using the above code, we calculate a semi-cosmological model as follows.
We consider an isolated sphere on which small-scale fluctuations
corresponding to a CDM power spectrum are superimposed.
Here, we use Bertschinger's software COSMICS (Bertschinger 1995)
in generating initial fluctuations.
To incorporate the effects of fluctuations with longer wavelengths,
the density of the sphere has been enhanced and a rigid rotation
corresponding to a spin parameter, $\lambda$, has been added.
The initial condition of this model is determined
by four parameters $\lambda$, $M_{\rm tot}$,
$\sigma_{\rm 8,in}$, and $z_{\rm c}$:
the spin parameter is defined by
\begin{equation}
 \lambda \equiv \frac{J|E|^{1/2}}{G M_{\rm tot}^{5/2}},
\end{equation}
where $J$ is the total angular momentum of the system,
$E$ is the binding energy when the system turns around,
and $M_{\rm tot}$ is the total mass of this sphere, which is
composed of dark matter and gas;
$\sigma_{\rm 8,in}$ is an rms mass fluctuation in a sphere of radius
$8\ h^{-1}$ Mpc, which normalizes the amplitude of the CDM power spectrum;
$z_c$ is the expected collapse redshift.
 If the top-hat density perturbation has an amplitude of
$\delta_i$ at the initial redshift, $z_i$, we have 
$z_{\rm c} = 0.36 \delta_i (1+z_i)-1$ (e.g., Padmanabhan 1993).
Thus, when $z_{\rm c}$ is given, $\delta_i$ at $z_i$ is determined.

Our simulations assume a flat universe ($\Omega = 1$)
with a baryon fraction of $\Omega_b = 0.1$
and a Hubble constant of $H_0 = 50\ {\rm km\ s^{-1}\ Mpc^{-1}}$.
In this paper, we focus on the evolution
of a top-hat density perturbation with
a spin parameter $\lambda=0.02$.
This is close to the minimum value of the spin parameter possible
in a CDM universe, according to the results of the N-body simulation
(Barnes, Efstathiou 1987; Warren et al.\ 1992).
Other parameters are $M_{\rm tot}=8 \times 10^{11}$ ${M_\odot}$,
$\sigma_{\rm 8,in} = 0.5$, and  $z_{\rm c}=2$.
Then, the radius of the proto-galactic sphere  is $32.17\ {\rm kpc}$
(in real space) at the initial redshift of $z_i = 40$.

 We carry out two simulations with the different
numbers of particles, ${N_{\rm p}}$.
One simulation has ${N_{\rm p}}=9171$ (Model 1) and another
has ${N_{\rm p}}=3071$ (Model 2). In Model 1,
the initial redshift is set to be $z_i=40$, while in Model 2,
 $z_i=25$.  The numerical simulation
should start from a redshift at which small scale perturbations are
weak enough. The simulation with large $N_{\rm p}$ can include
smaller-scale perturbations than that with small $N_{\rm p}$,
and the perturbations on a smaller scale have a larger amplitude
than large-scale perturbations in the CDM power spectrum.
This is why the initial redshift of Model 1 is taken to be higher.
In Model 1 (Model 2), the masses of gas, star, and dark matter particles
are $8.72 \times 10^6\ (2.61 \times 10^7)$,
$8.72 \times 10^6\ (2.61 \times 10^7)$, and
$7.85 \times 10^7\ (2.34 \times 10^8)\ {M_\odot}$, respectively, and
the softening lengths of gas, star, and dark matter particles
are $1.20\ (1.73)$, $1.20\ (1.73)$, and $2.50\ (3.61)\ {\rm kpc}$,
respectively.

\section{ Results }

 We simulate the evolution of the galaxy based on the two models from
the redshift $z_i$ to the present, $z=0$.
The angular momentum of the whole system is conserved with sufficient
accuracy: the error is less than 0.3 (1) $\%$ in Model 1 (Model 2).
Figure 1 shows the evolution of the system in Model 1.
These panels show the projection
of the particles to the $x$--$z$ plane, where we take the $z$-axis
to be the initial rotational axis.
 At redshift of $z=4.88$, the system is expanding with the Hubble Flow.
The amplitude of the initial top-hat density perturbation is chosen so that
the system should turn around at $z=3.71$.
 At $z=3.54$, small clumps are formed
and star formation begins in these clumps.
These clumps have merged and the whole system has collapsed 
at $z=1.89$. The stars are formed in a burst and
gas particles diminish, changing into star particles.
At $z=0.84$, the system has already relaxed and is evolving almost passively.
\begin{figure*}
\vspace{10cm}
\footnotesize Fig.\ 1.\ Time evolution of the system in Model 1.
The upper, middle, and lower panels show
the distributions of the dark matter, the gas,
and the stars, respectively. Each panel
measures 200 kpc across and shows
the $x$--$z$ projection of the particles,
 where we set the $z$-axis to be the initial rotational axis.
\end{figure*} 

 The history of the SFR is shown in figure 2.
 Figure 2 also shows the SFR in Model 2.
In spite of the difference in the number of particles,
the two simulations yielded similar results.
 From figure 2, we see that the major star formation
begins due to the collapse of whole system
and continues for about 1 Gyr with a rate of
about $50\ {M_\odot\ {\rm yr}^{-1}}$.
 At the age of universe  $t=4\ {\rm Gyr}\ (z\sim1.2)$,
the star formation is almost quenched
owing to the exhaustion of the gas.
\begin{figure*}
 \begin{center}
   \leavevmode  
   \epsfxsize=160mm
   \epsfbox{fig2.eps}
 \end{center}
\footnotesize Fig.\ 2.\  Star formation rate in ${M_\odot\ {\rm yr}^{-1}}$
as a function of the age of universe
in Model 1 (left) and Model 2 (right), respectively.
\end{figure*}

 The distribution of star particles at $z=0$
is shown in figure 3.  Figure 3 shows the $x$--$y$ and
$x$--$z$ projection. Both projections look quite similar.
The final stellar system seems to be
nearly spherical due to the low initial spin,
in contrast to the results of Katz (1992)
who made the disk system
contracted along the rotational axis
due to the high initial spin.
\begin{figure*}
 \begin{center} 
   \leavevmode  
   \epsfxsize=160mm
   \epsfbox{fig3.eps}
 \end{center}
\footnotesize Fig.\ 3.\  The particle distribution of stars at $z=0$.
The left (right) panel shows the $x$--$y$ ($x$--$z$) projection,
where the $z$-axis is the initial rotational axis.
Each panel measures 100 kpc across.
\end{figure*}
\begin{figure*}
 \begin{center}
   \leavevmode  
   \epsfxsize=160mm
   \epsfbox{fig4.eps}
 \end{center}
\footnotesize Fig.\ 4.\  The surface mass density profiles in Model 1 (left)
and Model 2 (right), respectively.
 The solid line shows the de Vaucouleurs law given by equation (24)
with parameters being listed in table 1.
\end{figure*}

 Figure 4 shows the surface mass density profile
of the final stellar system at $z=0$ in Model 1 and Model 2.
These surface densities are obtained for the cylindrical shells
of various radii in the $x$--$y$ projection. We have confirmed
that these profiles do not change in other projections.
The surface density profile in figure 4 is in good agreement with
the de Vaucouleurs law,
\begin{equation}
 \Sigma = \Sigma_{\rm e} \exp \left\{-7.67\left[(R/R_{\rm h,m})^{1/4}-1\right]
\right\},
\label{deVeq} 
\end{equation}
where $R_{\rm h,m}$ is the radius containing half the total mass and
$\Sigma_{\rm e}$ is the mass density at $R_{\rm h,m}$.
These parameters are summarized in table 1.
Here, $R_{\rm h,m}$ is comparable to the softening length
especially in Model 2.
In order to examine the effect of the softening length
of star particles, $\varepsilon_{\rm s}$, which equals that of gas particles,
we also carry out five simulations ($N_{\rm p}=3071$) with
the softening length $\varepsilon_{\rm s}=$
0.25$\varepsilon_{\rm s,0}$, 0.5$\varepsilon_{\rm s,0}$,
0.75$\varepsilon_{\rm s,0}$,
1.25$\varepsilon_{\rm s,0}$, and 1.5$\varepsilon_{\rm s,0}$,
where $\varepsilon_{\rm s,0}=1.73$ kpc denotes
the standard value used in Model 2.
Figure 5 shows $R_{\rm h,m}$ at $z=0$ as a function of
$\varepsilon_{\rm s}$.
It is clear that $\varepsilon_{\rm s,0}$ is an appropriate value,
because $R_{\rm h,m}$ does not depend on the softening length
around $\varepsilon_{\rm s,0}$.
It is worth noting that, in the range
$\varepsilon_{\rm s}<0.5\varepsilon_{\rm s,0}$,
the softening length is so small that
the artificial two-body relaxation causes the rise of $R_{\rm h,m}$.
Figure 6 shows this effect clearly.
In the case $\varepsilon_{\rm s}=\varepsilon_{\rm s,0}$,
$R_{\rm h,m}$ does not change from $z=0.58$,
at which the system has already relaxed,
to the present epoch $z=0$.
In the case $\varepsilon_{\rm s}=0.25 \varepsilon_{\rm s,0}$,
however, the central concentration is diluted
and $R_{\rm h,m}$ goes up at $z=0$.
Thus, the resulting value of $R_{\rm h,m}$ appears to be trustful
only over a limited range of $\varepsilon_{\rm s}$, which includes
$\varepsilon_{\rm s,0}$ adopted in Model 2.
\begin{figure}
 \begin{center}
   \leavevmode  
   \epsfxsize=80mm
   \epsfbox{fig5.eps}
 \end{center}
\footnotesize Fig.\ 5.\ Half-mass radius as a function of the softening length
of the star particles in the simulations with ${N_{\rm p}}=3071$.
A filled circle corresponds to the softening length
in Model 2 ($\varepsilon_{\rm s,0}=1.73$ kpc).
\end{figure}
\begin{figure}
 \begin{center}
   \leavevmode  
   \epsfxsize=80mm
   \epsfbox{fig6.eps}
 \end{center}
\footnotesize Fig.\ 6.\ The surface mass density profiles
at $z=0.58$ (short-dashed lines)
and at $z=0$ (solid lines), respectively.
The thick and normal lines correspond to the case of
$\varepsilon_{\rm s}=\varepsilon_{\rm s,0}$ and the case of
$\varepsilon_{\rm s}=0.25 \varepsilon_{\rm s,0}$,
respectively. The arrows indicate the position of the half-mass
radius.
\end{figure}
 
 It is known that a violent relaxation for sufficiently deep central
potentials reproduces the de Vaucouleurs law
(Hjorth, Madsen 1991 and references therein).
The deep potential causes the strong scattering
which tightly bound particles experience near the center, and
leads to the outer envelope of the de Vaucouleurs profile.
In dissipationless collapse, a cold collapse is preferred
to yield a deep enough central potential (e.g., van Albada 1982).
However, the dissipation is
a crucial component of galaxy formation.
Carlberg et al. (1986) found
that dissipation, leading to a deep central potential,
could help get a good de Vaucouleurs profile.
We can infer that our numerical simulations also result in
the de Vaucouleurs profile owing to the collapse combined with
the dissipation.

 In Model 1, we analyze the kinematic properties,
the rotation velocity and the velocity dispersion.
The maximum rotation velocity, $V_{\rm m}$, is 25 km/s
at the radius of 9 kpc where the rotation curve flattens out,
and the mean velocity dispersion, $\overline{\sigma}$,
within the half-mass radius is 116 km/s. 
$V_{\rm m}/\overline{\sigma}$ is a important value
characterizing the kinematic properties of elliptical galaxies.
$V_{\rm m}/\overline{\sigma}=0.2$ means
that this system is slowly rotating
and supported by the velocity dispersion due to the low initial spin.
This value resembles that of bright elliptical
galaxies (e.g., Davies et al. 1983). 

 Our simulation includes the chemical evolution process.
As a result, each star particle has 
its own age and metallicity. From this information,
we can obtain the SED (Spectral Energy Distribution)
using the SSPs (Single Stellar Populations) in Kodama, Arimoto (1997).
Consequently, we can analyze distributions of the luminosity and color.
Figure 7 shows the surface brightness
(${\rm L_{\odot,{\it V}}\ pc^{-2}}$) in $V$ band
as a function of the radius in Model 1 and Model 2.
The surface brightness profile also matches the de Vaucouleurs
law very well, using the half-light radius $R_{\rm h,l}$ and
the luminosity $I_{\rm e}$ at $R_{\rm h,l}$ given in table 1.
\begin{figure*}
 \begin{center}
   \leavevmode  
   \epsfxsize=160mm
   \epsfbox{fig7.eps}
 \end{center}
\footnotesize Fig.\ 7.\  The surface brightness profiles in Model 1 (left)
and Model 2 (right), respectively.
 The solid line shows the de Vaucouleurs law
with parameters being listed in table 1.
\end{figure*}

\begin{table*}[t]
\begin{center}
 Table~1.\hspace{4pt} de Vaucouleurs law parameters
 and photometric properties.\\
\end{center}
\vspace{6pt}
\begin{tabular*}{\textwidth}{@{\hspace{\tabcolsep}
\extracolsep{\fill}}cccccccc}
\hline\hline
& ${N_{\rm p}}$ & $\Sigma_{\rm e}$
 & $R_{\rm h,m}$ & $I_{\rm e}$
 & $R_{\rm h,l}$ & $M_V$ & $V-K$ \\
&  & $({M_\odot\ {\rm pc}^{-2}})$
 & $({\rm kpc})$ & $({\rm L_{\odot,{\it V}}\ pc^{-2}})$
 & $({\rm kpc})$ & & $(<10{\rm kpc})$
\\[4pt]\hline
Model 1 & 9171 & 250 & 3.54 & 23 & 4.49 & $-20.34$ & 3.164\\
Model 2 & 3071 & 250 & 3.99 & 23 & 4.99 & $-20.34$ & 3.166\\ 
\hline
\end{tabular*}
\end{table*}
 
 Based on the luminosity of each star particle,
we also obtain the radial profile of 
the metallicity and color weighted by the luminosity.
Figure 8 shows the projected
metallicity and the $V-K$ color profile in Model 1.
The metallicity has a gradient of
$\Delta \log Z/\Delta \log r \sim -0.3$, which is consistent with
the value $\Delta \log Z/\Delta \log r = -0.2\pm 0.1$ observed
by Davies et al.\ (1993). The $V-K$ color also has a similar amount
of gradient which is caused mainly by the spatial change of metallicity.
\begin{figure*}
 \begin{center}
   \leavevmode  
   \epsfxsize=160mm
   \epsfbox{fig8.eps}
 \end{center}
\footnotesize Fig.\ 8.\ The projected metallicity (left)
and the $V-K$ color (right)
versus the radius in Model 1.
In the left panel, the solid line has the
gradient of $\Delta \log Z/\Delta \log r=-0.3$.
\end{figure*}

 In table 1, we summarize the photometric properties
of this stellar system at $z=0$.
The eighth column in table 1
gives the $V-K$ color within a 10 kpc aperture, which is the same
aperture as adopted in the observation by Bower et al. (1992),
who investigated the color--magnitude relation of
early-type galaxies in the Coma and Virgo clusters.
 The half-light radius, the metallicity gradient, and the $V-K$ color
obtained in the simulation agree with those values observed for
a $M_V\sim-20$ elliptical galaxy (e.g.,
Faber et al.\ 1989; Peletier et al. 1990; Bender et al.\ 1992;
Bower et al.\ 1992; Davies et al.\ 1993).
 There is no significant difference between the two simulations
which have different $N_{\rm p}$.
Therefore, these simulations 
have a sufficiently large particle number and, hence,
sufficient mass resolution to reproduce the main properties of
the elliptical galaxy with enough accuracy.

\section{Summary and Discussion}

 We simulated the galaxy formation from a top-hat
density perturbation with a small spin parameter, $\lambda=0.02$.
The stellar system formed in the simulation
is spherical and has a mass distribution
which is in good agreement with the de Vaucouleurs law.
This system was formed through
turning around from the Hubble expansion
and subsequent merging of small clumps
originating in initial CDM density perturbations.
We also obtained photometric properties
by simulating star formation process
which follows chemical evolution.
These properties were found to be similar to the observed properties
of elliptical galaxies. The results of our simulation
clearly show that, in a CDM universe,
the proto-galaxy which has a spin-parameter as small as 0.02
evolves into an elliptical galaxy.

 Using three-dimensional SPH numerical simulation,
Mori et al.\ (1999) investigated elliptical galaxy formation.
They assumed a virialized proto-galactic gaseous cloud as
the initial condition. As Katz \& Gunn (1991) showed, however,
the gas never gets heated to the virial temperature
in the hierarchical galaxy formation which was adopted in our simulations.
The present simulations thus improve on their non-cosmological simulations.

In our numerical simulation,
the luminosity profile is in good agreement with
the de Vauclouleurs law as well as the mass distribution,
in spite of the difference between the half mass and light radii.
Figure 9 shows the projected stellar mass-to-luminosity ratio in $V$ band
as a function of the radius in Model 1.
Due to the metallicity gradient, the mass-to-luminosity ratio
decreases as the radius increases. This gradient
of mass-to-luminosity ratio leads to
the half-light radius larger than the half-mass radius.
However, the de Vauclouleurs law is held for the luminosity,
if the mass-to-luminosity ratio varies with respect to the radius
as follows,
\begin{equation}
 \frac{\Sigma(r)}{I(r)}
 = \frac{\Sigma_{\rm e}}{I_{\rm e}}
 \exp \left[-7.67
   \left(R_{\rm h,m}^{-1/4}-R_{\rm l,m}^{-1/4} \right) r^{1/4} \right].
\label{deVeq} 
\end{equation}
In figure 9, the solid line corresponds to this equation
with parameters of table 1. The simulation results match this line.
Thus, this gradient of mass-to-luminosity ratio
does not make the luminosity profile
fail to reproduce the de Vaucouleurs law.
\begin{figure}
 \begin{center}
   \leavevmode  
   \epsfxsize=80mm
   \epsfbox{fig9.eps}
 \end{center}
\footnotesize Fig.\ 9.\  The projected stellar mass-to-luminosity ratio
in $V$ band as a function of the radius in Model 1.
The solid line corresponds to equation (25) with parameters
of table 1.
\end{figure}

 It is notable that we obtain the projected metallicity gradient
which is consistent with the observed one for elliptical galaxies.
This gradient is caused by different star formation history
at different sites.
Figure 10 shows the time variations of SFR
in the regions of inside and outside the half-light radius in Model 1.
It is clear that the star formation holds on longer time
in the inner region than in the outer region.
Since the residual gas is polluted by the past star formation,
metal-rich stars are formed in the central region
where the duration of the star formation is long.
\begin{figure}
 \begin{center}
   \leavevmode  
   \epsfxsize=80mm
   \epsfbox{fig10.eps}
 \end{center}
\footnotesize Fig.\ 10.\ The history of the star formation rate (SFR)
 in different regions in Model 1.
 The solid line is the SFR within the half-light radius,
while the dotted line is that outside the half-light radius. 
\end{figure}

 Since the cooling of the gas overcame the feedback from massive stars,
our simulations did not lead to outflow of the gas in the form of
a galactic wind, which is often postulated to explain
photometric properties of the elliptical galaxy population
(e.g., Larson 1974b, Arimoto, Yoshii 1987).
Although our numerical model ignored some important feedback processes
like the stellar wind and the Type Ia supernova,
the system which we simulated might be too massive to allow a galactic wind.
To confirm this conjecture, we need to simulate the systems of
different masses. The ultimate goal of such
an extensive attempt is to reproduce the color--magnitude relation
of elliptical galaxies (e.g., Bower et al.\ 1992).
A forthcoming paper will discuss the properties of model elliptical
galaxies as a function of galaxy mass.

\vspace{1pc}\par
D.K. would like to thank M. Noguchi for valuable discussion,
and the editor, S. Mineshige, for correcting the draft of this paper.
D.K. also acknowledges the Astronomical institute, Tohoku University
and the Astronomical Data Analysis Center of the National Astronomical
Observatory, Japan
where the numerical computations presented in this paper were performed.
This work was supported in part by
the Japan Atomic Energy Research Institute.
 
\section*{References}
\re
 Arimoto N., Yoshii Y.\ 1987, A\&A 173, 23
\re
 Barnes J.E., Hut P.\ 1986, Nature 324, 446
\re
 Barnes J., Efstathiou G.\ 1987, ApJ 319, 575
\re
 Bender R., Burstein D., Faber S.M.\ 1992, ApJ 399, 462
\re
 Bertschinger E.\ 1995 preprint (astro-ph/9506070)
\re
 Blumenthal G.R., Faber S.M., Primack J.R. Rees M.J.\ 1984,
Nature, 311, 517
\re
 Bower R.G., Lucey J.R., Ellis R.S. 1992, MNRAS 254, 601
\re
 B\"ohringer H., Hensler G.\ 1989, A\&A 215, 147
\re
 Carlberg R.G.\ 1984a, ApJ 286, 403
\re
 Carlberg R.G.\ 1984b, ApJ 286, 416
\re
 Carlberg R.G., Lake G., Norman C.A.\ 1986, ApJ 300, L1
\re
 Carraro G., Lia C., Chiosi C.\ 1998, MNRAS 297, 1021
\re
 Dalgarno A., McCray R.A.\ 1972, ARA\&A 10, 375
\re
 Davies R.L., Efstathiou G., Fall S.M., Illingworth G.,
Schechter P.L.\ 1983, ApJ 266, 41
\re
 Davies R.L., Sadler E.M., Peletier R.F.\ 1993, MNRAS 262, 650
\re
 Eisenstein D.J., Loeb A.\ 1995, ApJ, 439, 520
\re
 Faber S.M., Wegner G., Burstein D., Davies R.L., Dressler A.,
Lynden-Bell D., Terlevich R.J.\ 1989, ApJS, 69, 763
\re
 Gingold R.A., Monaghan J.J.\ 1977, MNRAS 181, 375
\re
 Heavens A., Peacock J. 1988, MNRAS 232, 339
\re
 Hernquist L., Katz N.\ 1989, ApJS 70, 419
\re
 Hjorth J., Madsen J.\ 1991, MNRAS 253, 703
\re
 Katz N.\ 1992, ApJ 391, 502
\re
 Katz N., Gunn J.E.\ 1991, ApJ 377, 365
\re
 Katz N., Weinberg D.H., Hernquist L.\ 1996, ApJS 105, 19
\re
 Kawata D., Hanami H.\ 1998, PASJ 50, 547
\re
 Kodama T., Arimoto N.\ 1997, A\&A 320, 41
\re
 Larson R.B.\ 1969, MNRAS 145, 405
\re
 Larson R.B.\ 1974a, MNRAS 166, 585
\re
 Larson R.B.\ 1974b, MNRAS 169, 229
\re
 Larson R.B.\ 1975, MNRAS 173, 671
\re
 Larson R.B.\ 1976, MNRAS 176, 31
\re
 Lucy L.B.\ 1977, AJ 82, 1013
\re
 Maeder A.\ 1987, A\&A 173, 247
\re
 Makino J.\ 1990, J.\ Comp.\ Phys 87, 148
\re
 Makino J.\ 1991, PASJ 43, 859
\re
 Mori M., Yoshii Y., Nomoto K.\ 1999, ApJ 511, 585
\re
 Navarro J.F.,\ Steinmetz M.\ 1997, ApJ 478, 13
\re
 Navarro J.F., White S.D.M.\ 1993, MNRAS 265, 271
\re
 Padmanabhan T.\ 1993, Structure formation in the universe
(Cambridge Univ. Press, Cambridge) ch8
\re
 Pealzner S., Gibbon P. 1996, MANY-BODY TREE METHODS IN PHYSICS
(Cambridge Univ. Press, Cambridge) ch2
\re
 Peebles P.J.E.\ 1971, A\&A 11, 377
\re
 Peletier R.F., Valentijn E.A., Jameson R.F.\ 1990, A\&A 233, 62
\re
 Steinmetz M.\ 1996, MNRAS 278, 1005
\re
 Steinmetz M., Bartelmann M.\ 1995, MNRAS 272, 570
\re
 Steinmetz M., M\"uller E.\ 1994, A\&A 281, L97
\re
 Steinmetz M., M\"uller E.\ 1995, MNRAS 276, 549
\re
 Theis Ch., Burkert A., Hensler G.\ 1992, A\&A 265, 465
\re
 van Albada T.S.\ 1982, MNRAS 201, 939
\re
 Warren M., Quinn P.J., Salmon J.K., Zurek W.H., 1992, ApJ 399, 405

\end{document}